\documentclass[twocolumn,trackchanges]{aastex7} 
\usepackage{amsmath}
\usepackage{soul}

\begin{document}

\title{The Stellar Mass and Age Distributions of Star-Forming Clumps at $0.5 < z < 5$ in JWST CANUCS: Implications for Clump Formation and Destruction}

\author[0000-0003-0780-9526]{Visal Sok}
\affiliation{Department of Physics and Astronomy, York University, 4700 Keele Street, Toronto, ON, M3J 1P3, Canada}
\email[show]{sokvisal@yorku.ca}  

\author[0000-0002-9330-9108]{Adam Muzzin} 
\affiliation{Department of Physics and Astronomy, York University, 4700 Keele Street, Toronto, ON, M3J 1P3, Canada}
\email{muzzinad@yorku.ca}  

\author[0000-0002-3503-8899]{Vivian Yun Yan Tan}
\affiliation{Department of Physics and Astronomy, York University, 4700 Keele Street, Toronto, ON, M3J 1P3, Canada}
\email{tanvivia@yorku.ca}

\author[0000-0003-3983-5438]{Yoshihisa Asada}
\affiliation{Dunlap Institute for Astronomy and Astrophysics, 50 St. George Street, Toronto, Ontario, M5S 3H4, Canada}
\affiliation{Waseda Research Institute for Science and Engineering, Faculty of Science and Engineering, Waseda University,\\ 3-4-1 Okubo, Shinjuku, Tokyo 169-8555, Japan}
\email{asada@kusastro.kyoto-u.ac.jp}

\author[0000-0001-5984-0395]{Maru\v{s}a Brada\v{c}}
\affiliation{Faculty of Mathematics and Physics, Jadranska ulica 19, SI-1000 Ljubljana, Slovenia}
\affiliation{Department of Physics and Astronomy, University of California Davis, 1 Shields Avenue, Davis, CA 95616, USA}
\email{marusa.bradac@fmf.uni-lj.si}

\author[0000-0001-8489-2349]{Vince Estrada-Carpenter}
\affiliation{School of Earth and Space Exploration, Arizona State University, Tempe, AZ 85287, USA}
\affiliation{Beus Center for Cosmic Foundations, Arizona State University, Tempe, AZ 85287, USA}
\affiliation{Department of Astronomy and Physics and Institute for Computational Astrophysics, Saint Mary's University, 923 Robie Street, Halifax, Nova Scotia B3H 3C3, Canada}
\email{vestrad9@asu.edu}

\author[0000-0001-9298-3523]{Kartheik G. Iyer}
\affiliation{Columbia Astrophysics Laboratory, Columbia University, 550 West 120th Street, New York, NY 10027, USA}
\affiliation{Center for Computational Astrophysics, Flatiron Institute, 162 Fifth Avenue, New York, NY 10010, USA}
\email{kiyer@flatironinstitute.org}

\author[0000-0003-3243-9969]{Nicholas S. Martis}
\affiliation{Faculty of Mathematics and Physics, Jadranska ulica 19, SI-1000 Ljubljana, Slovenia}
\email{nicholas.martis@fmf.uni-lj.si}

\author{Ga\"el Noirot}
\affiliation{Space Telescope Science Institute, 3700 San Martin Drive, Baltimore, Maryland 21218, USA}
\email{gnoirot@stsci.edu}

\author[0000-0001-8830-2166]{Ghassan T. E. Sarrouh}
\affiliation{Department of Physics and Astronomy, York University, 4700 Keele Street, Toronto, ON, M3J 1P3, Canada}
\email{gsarrouh@yorku.ca}

\author[0000-0002-7712-7857]{Marcin Sawicki}
\affiliation{Department of Astronomy and Physics and Institute for Computational Astrophysics, Saint Mary's University, 923 Robie Street, Halifax, Nova Scotia B3H 3C3, Canada}
\email{marcin.sawicki@smu.ca}

\author[0000-0002-4201-7367]{Chris J. Willott}
\affiliation{National Research Council of Canada, Herzberg Astronomy \& Astrophysics Research Centre, 5071 West Saanich Road, Victoria, BC, V9E 2E7, Canada}
\email{chris.willott@nrc.ca}

\author[0009-0000-8716-7695]{Sunna Withers}
\affiliation{Department of Physics and Astronomy, York University, 4700 Keele Street, Toronto, ON, M3J 1P3, Canada}
\email{sunnaw@my.yorku.ca}

\author[0000-0002-3503-8899]{Samantha C. Berek}
\affiliation{David A. Dunlap Department of Astronomy and Astrophysics, University of Toronto, 50 St. George Street, Toronto, Ontario, M5S 3H4, Canada}
\affiliation{Dunlap Institute for Astronomy and Astrophysics, 50 St. George Street, Toronto, Ontario, M5S 3H4, Canada}
\affiliation{Data Sciences Institute, University of Toronto, 17th Floor, Ontario Power Building, 700 University Ave, Toronto, ON M5G 1Z5, Canada}
\email{sam.berek@mail.utoronto.ca}

\author[0009-0009-2307-2350]{Katherine Myers}
\affiliation{Department of Physics and Astronomy, York University, 4700 Keele Street, Toronto, ON, M3J 1P3, Canada}
\email{kjmyers@yorku.ca}


\begin{abstract}

We investigate the resolved properties of star-forming clumps and their host galaxies at $0.5<z<5$ in the JWST CANUCS fields. We find that the fraction of clumpy galaxies peaks near $z\sim2$ for galaxies with masses of $\log(M_{g,*}/M_\odot)\geq10$. Galaxies with masses of $8.5 \leq \log(M_{g,*}/M_\odot) < 10$ show lower clumpy fractions with little redshift evolution. We identify and measure individual clump masses, finding that the aggregated clump stellar mass function (cSMF) follows a power-law slope of $\alpha = -2$ across all redshift bins, broadly consistent with \textit{in-situ} clump formation. However, when split by galaxy mass, the cSMF is found to be flatter ($\alpha\sim-1.6$) for massive galaxies and steeper ($\alpha\sim-2.3$) for lower mass galaxies, with little redshift evolution. We further explore how clump formation and disruption mechanisms shape the cSMF. In particular, we find that the cSMF slope is flatter with increasing inferred gas fraction in younger clump populations ($<300$ Myr), suggesting that gas availability favors the formation of more massive clumps. Alternatively, many high-redshift galaxies in our sample have disturbed morphologies and simulations show that clumps of \textit{ex-situ} origins are comparably more massive to \textit{in-situ} clumps, which can flatten the cSMF. We also examine clump evolution, finding the cSMF slope become flatter as clumps evolve and age. We interpret this as an indication of the long-term survivability of massive clumps, where feedback mechanisms preferentially disrupt low-mass clumps. Overall, the galaxy-mass dependent cSMF and age distribution point to a complex history for clumps, involving different and competing mechanisms for their formation and destruction. 

\end{abstract}




\section{Introduction} 

Star-forming galaxies at $z>1$ commonly exhibit irregular morphologies and often host multiple kiloparsec-scale, rest-frame UV bright star-forming clumps \citep[e.g.,][]{Elmegreen2005, Puech2010, ForsterSchreiber2011a, ForsterSchreiber2011b, Wisnioski2011, Wuyts2012, Wuyts2013, Tan2024}. Particularly, the fraction of clumpy galaxies is found to increase with redshift \citep[e.g.,][]{Murata2014, Guo2015, Huertas2020, Sok2022, Mercier2025}, with a peak around $z\sim2$, coincident with the peak of the cosmic star formation density \citep{Shibuya2016, Sattari2023, Vega2025}. With the recent launch of JWST, there is also a growing number of spectacularly gravitationally lensed sources observed after cosmic noon ($z\gtrsim5$) that host compact stellar structures \citep[e.g.,][]{Adamo2024, Mowla2024, Fujimoto2025, Vanzella2025}. While many of these objects are likely to be globular progenitors, it naturally raises important questions about how clumpy star formation occurs, what formation pathways dominate, and how these different mechanisms may evolve across cosmic time. 

The commonly suggested mechanisms for clump formation at $z\sim2$ include both \textit{in-situ} and \textit{ex-situ} processes. Clumps of \textit{in-situ} origins are typically linked to violent disk instability (VDI) as the result of sustained cosmological gas accretion, with gravitational fragmentation governed by the Toomre $Q$ instability parameter \citep[e.g.,][]{Toomre1964, Agertz2009, Dekel2009_clumpy_disks, Dekel2009_cold_stream, Ceverino2010, Cacciato2012}. This is motivated by observations of very gas-rich galaxies at higher redshifts \citep[e.g.,][]{Daddi2010, Tacconi2010, Wang2022}, with well-ordered rotation (e.g., \citealt{ForsterSchreiber2018, Wisnioski2019}). The fragmentation of clumps in gas-rich disks is also well-studied in simulations of galaxies in both isolation and cosmological context \citep[e.g.,][]{Ceverino2016, Fensch2021, Renaud2024}. Alternatively, clumps of \textit{ex-situ} origins are attributed to gas compression and fragmentation triggered by galaxy interactions \citep[e.g.,][]{Puech2010, Calabro2019, Estrada2024}. These clumps could be remnants of accreted satellites as shown in simulations \citep{Mandelker2017} and suggested in observations \citep{Ribeiro2017, Zanella2019}. Further supports for \textit{ex-situ} clumps are also found in \cite{Estrada2025}, where some clumps are found to have lower metallicity offsets compared to their host gas disks. These metallicity offsets are not expected if clumps are formed through disk instabilities, suggesting that these are perhaps satellites or even accreting clumps of near-pristine gas. 
Such metallicity offsets are also reported in \cite{Sok2025}, where clumpy galaxies are found to have lower metallicities compared to nonclumpy galaxies. 

Clump formation due to VDI has been previously argued to be more efficient near cosmic noon due to both the declining of cold gas accretion toward lower redshifts \citep[e.g.,][]{Daddi2022, Waterval2025} and the generally larger disks toward lower redshifts. Recent studies of galaxy morphologies at higher-$z$, however, suggest that VDI-driven clump formation may still be relevant at earlier epochs, given the relatively high fraction of disk-like morphologies observed out to $z \sim 7$ \citep[e.g.,][]{Ferreira2023, Kartaltepe2023}. 
At $z\gtrsim5$, both VDI and mergers are linked to lensed galaxies that exhibit clumpy morphologies as observed by JWST/NIRSpec IFU observations \citep[e.g.,][]{Fujimoto2025, Morishita2025}. Simulations also offer contrasting pictures. \cite{Nakazato2024} found that clump formation due to VDI did not occur until $z\sim6$ in their FirstLight simulation, with their clumps found at $5.5<z<9$ are associated with mergers. In contrast, hydrodynamical simulations in \cite{Mayer2025} showed clump formation due to fragmentation at $z>7$ is possible.


Larger, statistical samples of clumps in galaxies are needed to better constrain clump formation mechanisms. Particularly, the clump stellar mass distribution also provides insights into these processes. Detecting clumps and measuring their stellar masses accurately depend both on the spatial resolution and sensitivity of the observations. 
High-z star-forming clumps are generally observed to be larger and more massive, often estimated to be $10^6$-$10^9$ solar masses \citep[e.g.,][]{Kalita2025b_nir_clumps, Kalita2025a_csmf}, compared to their star-forming region counterparts in the local universe. While observations of gravitationally lensed galaxies have shown that clump masses are overestimated in those measured in unlensed galaxies \citep[e.g.,][]{Livermore2015, Cava2018, Mestric2022, Messa2024, Claeyssens2025}, the effect of degraded resolution is found to be negligible compared to the impact of observational sensitivity, which can overestimate masses by an order of magnitude \citep{Dessauges2017, Tamburello2017}. \cite{Dessauges2018} also showed that the slope of the clump stellar mass function is unaffected when limiting their sample to just blended clumps. Any inferred physics using the measured clump properties must therefore account for these sensitivity limits to ensure robust conclusions. 

Initial analyses on the redshift evolution of the clump stellar mass distribution at $z<5$ include the work of \cite{Dessauges2018} and \cite{Claeyssens2025}, where both studies found that clump formation is broadly consistent with being \textit{in-situ}. In this paper, we expand on these analyses by utilizing the deep, multi-wavelength imaging of the Canadian NIRISS Unbiased Cluster Survey (CANUCS) to constrain the clump properties of a larger sample of $\sim5000$ star-forming galaxies (SFGs) at $0.5<z<5$. In particular, we analyze the redshift evolution of the clump stellar mass function in different galaxy mass bins, with clump completeness considerations to mitigate sensitivity biases. We further investigate the relationship between the mass and age of clumps and their host galaxy properties within the context of the mechanisms for their formation and destruction. In order to obtain robust estimates of the stellar mass, age and star formation rate of clumps, we use images that are PSF-matched to the F444W filter. This gives an angular resolution of ${\sim}0.16^{\prime\prime}$, corresponding to an effective spatial resolution of ${\sim}1-1.3 ~\mathrm{kpc}$ for over our redshift range. In \S 
\ref{sec:data}, we briefly describe the CANUCS dataset, while \S \ref{sec:sample} covers our galaxy selection criteria for the analyses. In \S \ref{sec:analyses}, we describe our methodology, which includes spatially fitting the SED of galaxies, extracting clumps, and estimating our clump detection limits. We discuss the redshift evolution of the fraction of clumpy galaxies in \S \ref{sec:clumpyfrac}, as well as the clump properties and the clump stellar mass distribution in \S \ref{sec:clump_prop}. 

Throughout the paper, all magnitudes are expressed in the AB system \citep{Oke1983}. We adopt a $\Lambda$CDM cosmological model of the universe with $\Omega_\lambda$ = 0.7, $\Omega_\mathrm{M}$ = 0.3, and a Hubble constant of $H_0$ = 70 km/s/Mpc.

\section{Data} \label{sec:data}

This paper makes use of deep and extensive photometric datasets from the Canadian NIRISS Unbiased Cluster Survey (CANUCS, \citealt{Willott2022}). A detailed overview of the first data release is presented by \cite{Sarrouh2025}. Briefly, the full datasets for CANUCS include HST, JWST/NIRCam and JWST/NIRISS photometry, along with spectroscopy from JWST/NIRISS and JWST/NIRSpec of five strong lensing clusters; Abell 370 ($z=0.375$, hereafter A370, \citealt{Lotz2017}), MACS J0416.1-2403 ($z=0.395$, hereafter MACS0416, \citealt{Lotz2017}), MACS J0417.5-1154 ($z=0.443$, hereafter MACS0417, \citealt{Postman2012}), MACS J1149.5+2223 ($z=0.543$, hereafter MACS1149, \citealt{Lotz2017}), and MACS J1423.8+2404 ($z=0.545$, hereafter MACS1423, \citealt{Salmon2020}). The primary cluster fields consist of imaging from NIRISS, including filters F090WN, F115WN, F150WN, and F200WN (where the suffix ``N'' denoting the NIRISS filter designation), with NIRCam filters F090W, F115W, F150W, F200W, F277W, F356W, F410M, and F444W. Also included are supporting HST imaging, with filters F435W, F606W, F814W, F105W, F125W, F140W, and F160W. The typical $3\sigma$ magnitude limits are $rm_{\rm AB} \approx 29.5$ for NIRISS, $m_{\rm AB} \approx 29.5{-}30.0$ for NIRCam, and $m_{\rm AB} \approx 28.0{-}30.0$ for the HST/ACS and HST/WFC3 filters.

In addition to the primary cluster fields, there are a total of ten flanking fields, comprising of a NIRCam flanking (NCF) field and a NIRISS flanking (NSF) field that are observed by the NIRCam and NIRISS module on either side of each cluster field, respectively. In this work, we utilize the five NCF fields. Three of these five NCF fields (A370, MACS0416, MACS1149) include almost the full set of NIRCam imaging, while the other two NCF fields contain nine medium bands and four wide-band filters. In brief, the available filters for the NCF fields include wide-band filters F070W, F090W, F115W, F150W, F200W, F277W, F356W, and F444W, and medium-band filters F140M, F162M, F182M, F210M, F250M, F300M, F335M, F360M, F410M, F430M, F460M, and F480M. Similarly, the $3\sigma$ magnitude limits for the medium- and wide-band filters in the NCF fields are $m_{\rm AB} \approx 29.5{-}30.5$ for NIRCam, and $m_{\rm AB} \approx 27.5{-}30.0$ for the HST/ACS and HST/WFC3 filters. We refer the reader to \cite{Sarrouh2025} for the detailed list of available filters and limiting magnitude for each cluster and flanking field.

Details of the PSF-matching, object detection, photometry and catalog creation are described in \cite{Sarrouh2025}. Briefly, source detection is done based on a combined chi-mean detection image, with aperture photometry done using the \textsc{photutils} package \citep{Bradley2024}. Each cluster field is also affected by varying level of gravitational lensing, with one NIRCam module centered on the cluster and the other generally sampling a lower-magnification region. The lensing models for A370 are provided in \cite{Gledhill2024}, for MACS0416 in \cite{Rihtarsic2025}, MACS1149 in G. Rhitar\v{s}i\v{c} et al. (in prep), and MACS0417 and MACS1423 in G. Desprez et al. (in prep). The lensing models are available on the CANUCS website\footnote{\url{https://niriss.github.io/data.html}}. 

\section{Sample Selection} \label{sec:sample}

We make use of imaging and photometry products from Data Release 1 \citep{Sarrouh2025} to select the galaxy sample for this work. Objects with unreliable photometry, $\texttt{USE\_PHOT = False}$, are excluded. These include sources that are bright cluster galaxies, sources contaminated by the bright cluster galaxy subtraction, and those with Kron aperture overlapping masked and bad pixels. In addition, point-like objects are also removed to avoid stellar contamination, using the flag $\texttt{FLAG\_POINTSRC}$. 

\begin{figure}[!t]
    \centering
    \includegraphics[width=0.95\linewidth]{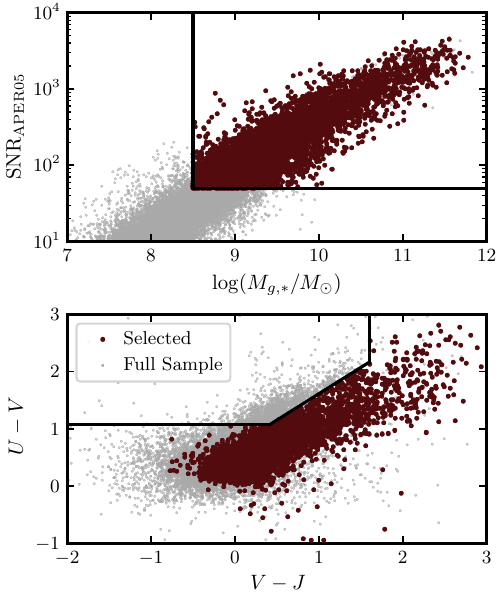}
    \caption{We selected galaxies based on their integrated SNR within a $0.5^{\prime\prime}$ aperture and stellar mass. Additionally, we used the $UVJ$ color selection to select star-forming galaxies (see text for details). Only ${\sim}20\%$ of all sources with $\texttt{USE\_PHOT = True}$ and $\texttt{FLAG\_POINTSRC = False}$ satisfies our selection criteria. It should also be noted that not all sources falling within each selection region necessarily satisfy all the selection criteria in this figure.}
    \label{fig:sample}
\end{figure}

Figure \ref{fig:sample} shows the main selection criteria of the sample. The selection criteria include cuts into stellar masses, redshift, integrate signal-to-noise ratio (SNR$_\mathrm{int}$, measured within an aperture diameter of $0.5^{\prime\prime}$ based on the F444W image), and rest-frame colors. 
The mass completeness of the CANUCS sample is determined to be 95\% complete down to stellar masses of $10^{8.2} ~\mathrm{M_\odot}$ at $z\sim5$ (Sarrouh et al. in prep.). We adopt a conservative cut at $10^{8.5} ~\mathrm{M_\odot}$, excluding lower-mass galaxies to enable a fair comparison of the clump stellar mass function across redshift. We exclude galaxies that have a SNR$_\mathrm{APER05}$ (SNR within an aperture of $0.5^{\prime\prime}$) less than 50 to ensure that each galaxy can be spatially binned using Voronoi tessellation, as described in \S \ref{sec:vorbin}, for resolved SED fitting. A lower redshift cut of $z=0.5$ ($z=0.65$ for MACS1149 and MACS1423 as these clusters are at a higher redshift) is also applied to exclude galaxies associated with the clusters, as our primary focus is on clump formation irrespective of environment. Due to the constraints on stellar masses and SNR$_\mathrm{APER05}$, galaxies at $z>5$ are also excluded as the number of galaxies that satisfy these criteria becomes limited.

Since clumps are regions of intense star formation, we select star-forming galaxies based on their rest-frame \textit{UVJ} colors. The \textit{UVJ} selection for SFGs at all redshift bins follows the prescription from \cite{Antwi2023}, and is defined by the following color cuts, 
\begin{equation}
    \begin{split}
        (U - V) &< 1.07 \\
        (V - J) &> 1.6 \\
        (U - V) &< (V - J) \times 0.92 + 0.69
    \end{split}
\end{equation}
\noindent These lines are calibrated for the CANUCS parent galaxy sample between $0.5<z<6$, where the slope and zero-point of the diagonal line is iteratively updated until a clear bimodality is determined between the red and blue sequence of galaxies (see \citealt{Tan2024}). Using a broad redshift interval here ensures that the calibration accounts for any redshift evolution in the zero-point. While the \textit{UVJ} color selection is known to lead to contamination of SFGs in the quiescent sample (e.g., \citealt{Schreiber2018}), it is still an efficient method to select different galaxy populations. We note other color selections can yield a more complete selection (e.g., \citealt{Noirot2022, Antwi2023}). However, since our analyses combine clump populations across many host galaxies, the results are not sensitive to the specific choice of color–color selection.

We also exclude galaxies with magnification factors of $\mu > 2$ as the majority of the galaxies in our sample is unlensed. This threshold removes ${\sim}600$ galaxies that are potentially subjected to significant differential magnification, but ensures that the analyses are performed on images with similar spatial resolution. We apply a correction to the derived properties for magnification by accounting for a factor of $\mu$ where applicable (see \S \ref{sec:retrieve_mass}). Lastly, we exclude highly edge-on galaxies with axial ratio $q<0.2$, as projection effects and increased dust attenuation in these systems can obscure clumps, leading to biases in clump detection and unreliable estimates of their physical properties. The axial ratio is determined by fitting a S\'ersic profile to each galaxy in the F444W filter with \textsc{GALFIT} \citep{Peng2002}. More details for the morphological parameter fitting is presented in Merchant et al. in prep. Finally,  the number of galaxies in our sample given these criteria is approximately 4500.


\section{Methodology} \label{sec:analyses}
This section describes the methodology for constructing resolved stellar mass density and rest-frame UV ($\lambda_{\rm rest} \approx 3600\,\text{\AA}$) surface brightness maps. Pixels are spatially binned to obtain the spectral energy distribution (SED). These SEDs are fitted to produce resolved maps of the stellar mass density and $U_\mathrm{rest}$ surface brightness maps. We adopt the recalibrated \textit{U} filter from \cite{Apellaniz2006}, which is based on the Johnson system, to retrieve $U_\mathrm{rest}$. We also discuss how clumps are identified within these $U_\mathrm{rest}$ brightness maps. Galaxies are first classified as clumpy and nonclumpy based on their $U_\mathrm{rest}$ brightness normalized profile, and individual clumps are identified in those galaxies which are classified as clumpy. Lastly, this section outlines how clump masses are retrieved from the two-dimensional stellar mass density maps and discuss our clump detection limits. Again, the following analyses are done using images that are PSF-matched to the F444W image, with pixel resolution of 40 mas.

\subsection{Spatially-resolved SED fitting}

\subsubsection{Voronoi Binning} \label{sec:vorbin}
Spatially-resolved SED fitting provides better estimates of stellar masses compared to unresolved fitting (e.g., \citealt{Sorba2015, Sorba2018}). However, rather than fitting the SED on a pixel-by-pixel basis, as individual pixels generally have low SNR, we first bin pixel together to acquire a minimum SNR. 


In this paper, the binning step is done based on Voronoi tessellation using \textsc{vorbin} \citep{Cappellari2003}, where pixels are binned until a minimum S/N threshold is reached in the F444W band. We adopt F444W for the binning as it provides the most reliable tracer of the underlying stellar mass.
The minimum SNR per Voronoi bin is defined by the integrated SNR of the galaxies in the F444W image (i.e., SNR$_\mathrm{APER05}$ as measured within a aperture diameter of 0.5$^{\prime\prime}$). We set the minimum SNR to be 5 and 10 for galaxies with $\mathrm{SNR_{APER05}}<500$ and $\mathrm{SNR_{APER05}}\geq500$, respectively. The choice for varying the minimum SNR on a galaxy to galaxy basis is because our sample consists of galaxies at a wide range of stellar masses and redshifts. A fixed threshold can otherwise yield fine binning in high-SNR galaxies and unnecessarily coarser binning in low-SNR galaxies.

\subsubsection{Inferring physical properties}
We determine the flux for each bin by summing the pixel values within that bin. Similar to \cite{Han2019}, the flux uncertainty for each bin is estimated as the quadrature sum of the statistical uncertainty and the systematic uncertainty. Here, the statistical uncertainty is the aperture flux uncertainty, while the systematic component accounts for potential biases during data reduction and calibration. Since each bin consists of multiple pixels, we assume that the flux uncertainty for a given bin is similar to that within a circular aperture encompassing the same area (see \citealt{Sarrouh2025} for estimations of the aperture flux uncertainty). We assume that the systematic uncertainty is just a multiplicative factor of the flux. A multiplicative value of 0.05 is used to account for systematic uncertainties in this analyses, similar to that proposed in \cite{Abdurrouf2021}.

We derive spatially resolved stellar mass and SED–based star formation rate surface densities for each galaxy by fitting the SED of every Voronoi bin with \textsc{Dense Basis} \citep{Iyer2019}. \textsc{Dense Basis} employs an amortized brute-force Bayesian approach to determine the star formation history non-parametrically, which has been shown help mitigate biases against outshining from younger stellar populations and lead to more robust measurements of the SFH (e.g., \citealt{Iyer2017, Leja2019, Lower2020}). We adopt a Chabrier initial mass function \citep{Chabrier2003} and Calzetti attenuation law \citep{Calzetti2000}. In later discussions, we also determine the age of clumps based on background subtracted aperture photometry. The clump age is defined to be the lookback time at which $75\%$ of its total stellar mass formed (denoted as $t_{75}$). We adopt $t_{75}$ to account for more recent star formation. To determine $t_{75}$, we sample the star formation history (SFH) posteriors, integrate each SFH to obtain the cumulative mass formed, and calculate the lookback time at which the cumulative mass fraction reaches 0.75. The adopted $t_{75}$ for each clump is taken as the median of the its posterior distribution. Additionally, we also spatially reconstruct the rest-frame UV surface brightness maps with \textsc{eazy-py} \citep{Brammer2008}, using the recalibrated \textit{U} filter from \cite{Apellaniz2006}. To this end, we fix the redshift of each SED to the redshift of its parent galaxy. 

Binning pixels together also effectively degrades the spatial resolution of our maps, especially at the outer region of galaxies where the SNR is lowest. This results in Voronoi bins that are generally larger at the outskirts. Large bins can also smear out clumps in the rest-frame UV brightness maps as binning is done based on the F444W image. To recover spatial resolution in the stellar mass density and $U_\mathrm{rest}$ brightness maps, we ``dezonify" the Voronoi bins by redistributing the binned values according to the flux in the JWST images, similar to the dezonification procedure applied in previous works (e.g., \citealt{Fernandes2013}). Specifically, for each pixel $i$ within a given Voronoi bin, the value is rescaled as,

\begin{equation}
Q^\prime_i = Q_\mathrm{bin} \frac{F_i}{F_\mathrm{bin}},
\end{equation}

\noindent where $Q_\mathrm{bin}$ is the SED-derived quantity for that bin (i.e., either mass or $U_\mathrm{rest}$ luminosity), $F_i$ is the flux at pixel $i$ (i.e., either using F444W or another image), and $F_\mathrm{bin}$ is the sum across all pixels in the bin. Explicitly, the F444W image is used to scale the stellar mass maps as F444W closely traces masses, while we use whichever image probes the rest-frame UV at a given redshift to scale the rest-frame UV surface brightness maps at that redshift. We remind the reader that all the images used in this paper are convolved to match the F444W PSF. This rescaling therefore preserves the total measured quantity for each bin while reintroducing spatial resolution at an angular resolution of ${\sim}0.16''$.

\begin{figure*}[!t]
    \centering
    \includegraphics[width=\linewidth]{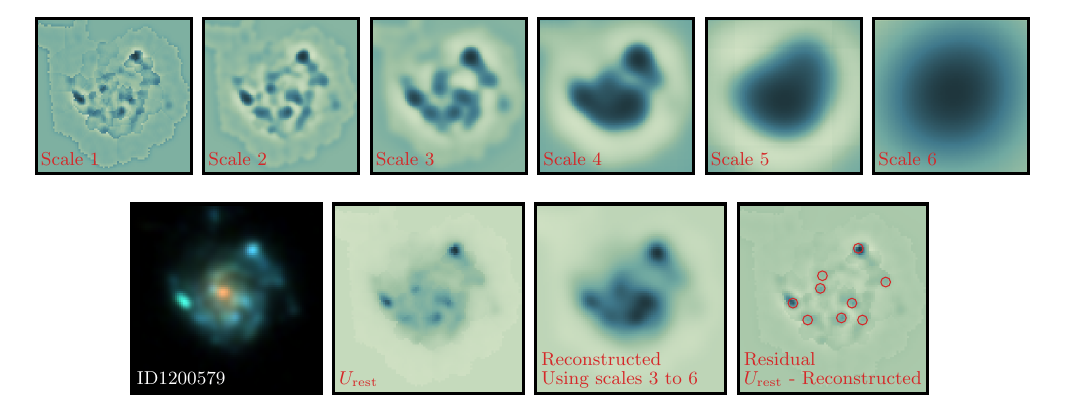}
    \caption{An illustration showing how clumps are identified with starlet wavelet transform. The rest-frame \textit{U} image is first decomposed into multiple spatial scales. The top row shows the resulting wavelet scales. A smooth model of the galaxy is then reconstructed by combining only the large-scale components (scales 3 and higher), while setting scales 1 and 2 to zero. The bottom row shows the composite color image, the rest-frame \textit{U} image, the reconstructed smooth image, and finally the high-contrast residual image obtained by subtracting the smooth model from the original. Clumps are identified using a peak finding algorithm on the high-contrast image (see text) and are marked by red circles with a diameter of $0.2^{\prime\prime}$. The cutouts shown here are $3.6^{\prime\prime}\times3.6^{\prime\prime}$.  }
    \label{fig:wavelet}
\end{figure*}

\subsection{Identifying clumps in the rest-frame UV maps} \label{sec:find_clumps}

We identify clumps in galaxies through two main steps. First, we select clumpy galaxy candidates using the normalized light profile method of \citet{Wuyts2012}. This method identifies broad regions of enhanced rest-frame UV emission, which include both extended and more localized clumpy star formation regions. Second, we apply a clump finding algorithm to find localized clumps in these clumpy candidates.

The normalized light profile is constructed based on the rest-frame UV surface brightness maps \citep[see][]{Wuyts2012, Sok2022, Sok2025}. We refer the reader to these papers for more details, however we briefly describe it here. Galaxies with smooth disk will have a normalized light profile that is brighter near its galactic center and tapers off with radius. As star-forming clumps are bright UV regions, superimposed on an otherwise smooth disk, clumpy regions will therefore appear as bumps in this one-dimensional normalized light profile. In order to make the normalized light profile for a galaxy, we first define two structural parameters of the galaxy; the axial ratio and the position angle. These elliptical parameters are inferred from the segmentation map of the galaxy. The apertures are then used to construct a curve of growth, with apertures centered on the mass-weighted center. We then define the half-light radius $R_e$ as the radius that contains half of the total light, and the effective surface brightness as the mean surface brightness within $R_e$. The normalized light profile is obtained by normalizing the distance of each pixel to the mass-weighted center of the galaxy by $R_e$, and the rest-frame UV surface brightness of each pixel by the effective surface brightness. Within this normalized parameter space, \cite{Wuyts2012} defined a region to identify clumpy pixels. Here, we classify a galaxy as a clumpy candidate if more than 5\% of its total rest-frame UV light arises from this clumpy region. The choice of 5\% is generally consistent with thresholds used in previous work to classify clumpy galaxies \citep[e.g.,][]{Guo2018, Sok2022, Sattari2023}.


\begin{figure*}[!t]
  \centering

  \includegraphics[width=\linewidth]{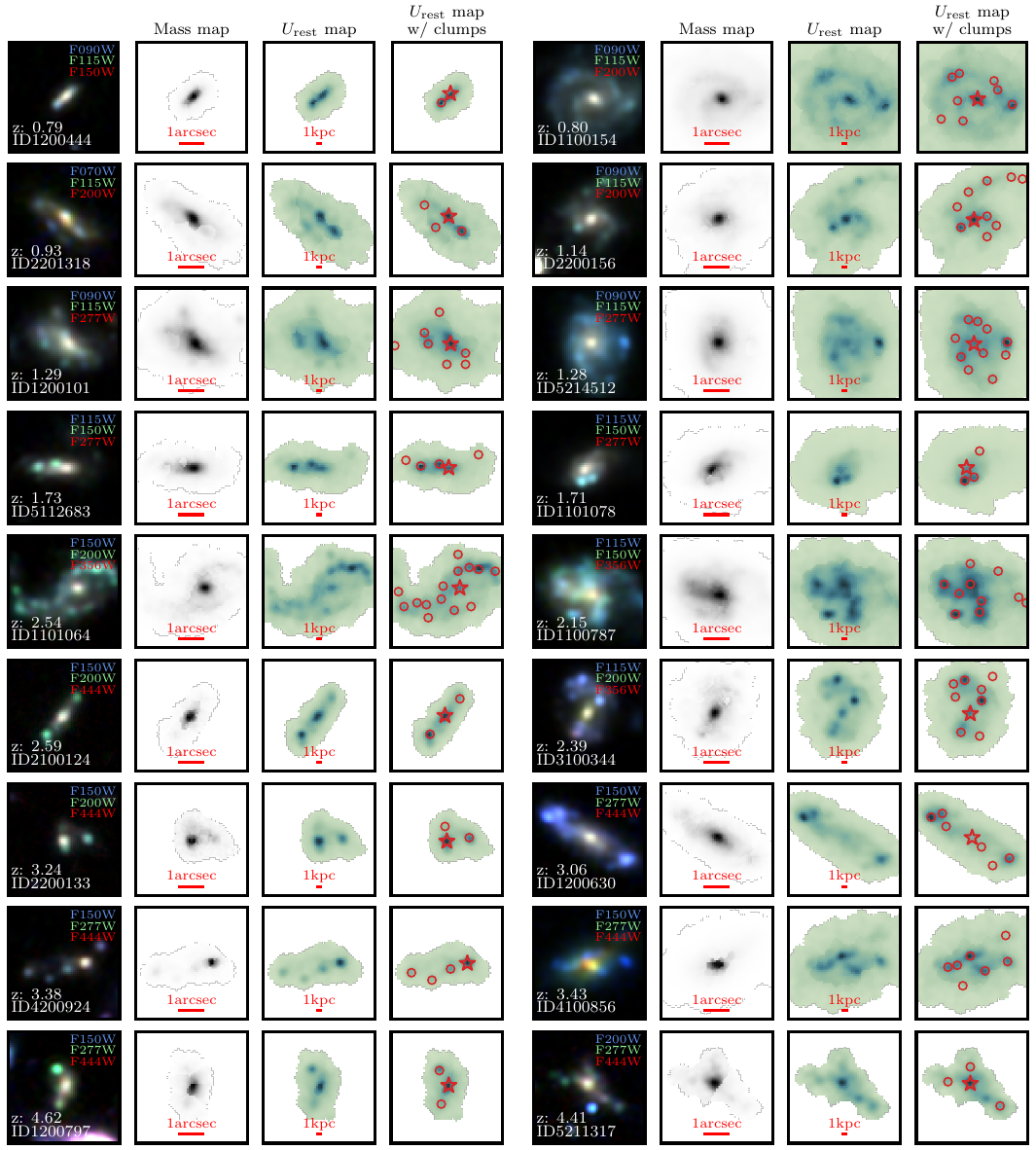}

  \caption{Examples of clumpy galaxies in the CANUCS fields. These galaxies are selected based on their clumpy morphology to illustrate our clump detection algorithm, and are ordered by redshift, with less massive galaxies ($\log(M_{g,*}/M_\odot) \lesssim 10 $) shown on the left and more massive galaxies ($\log(M_{g,*}/M_\odot) \gtrsim 10 $) on the right. In each panel, we show the composite RGB image, the stellar mass maps, the rest-frame UV map, and the same UV map with detected clumps (shown as the circle markers) and bulges (shown as the star markers).}
  \label{fig:clumpy_sfg}
\end{figure*}

The normalized light profiles only determine broad regions of enhanced rest-frame UV brightness, and will identify extended regions of star formation as clumpy. We identify star-forming clumps in these galaxies by using a clump finding algorithm similar to what is presented in \cite{Kalita2025b_nir_clumps}. We first subtract the large-scale structures from the rest-frame UV images to enhance small-scale structures such as clumps. This is achieved by decomposing the rest-frame UV images into six spatial scales using a wavelet-based approach, and reconstructing a smooth model of each imaging by setting scale one and two to zero. Specifically, we adopt the Isotropic Undecimated Wavelet Transform, or more commonly known as starlet transform, as it is well-suited for astronomical data analysis \citep{Starck2007}. The starlet transform has been implemented in the python package \textsc{scarlet} \citep{Melchior2018}.

Figure \ref{fig:wavelet} illustrates the steps to construct the high-contrast residual image. It should be noted that the third wavelet scale corresponds to an angular resolution of ${\sim}0.2^{\prime\prime}$. Subtracting wavelets of this scale and higher effectively removes any diffuse background emission and only leaves structures smaller than this scale. The resolution at scale 3 is determined by injecting a Gaussian profile with FWHM = 0.16$^{\prime\prime}$ into the image and analyzing its FWHM across wavelet scales. We apply a peak finding algorithm to find clump candidates using a threshold of $5\sigma$ and minimum separation of three pixels. To mitigate spurious detection, we further require that each detected clump has a flux above $3\sigma$ within a 0.2$^{\prime\prime}$-diameter aperture in the rest-frame UV. 

We note that the clump finding algorithm does not discriminate between star-forming clumps and bulge-like components. To remove contamination of bulge-like structures in the subsequent analyses, we assume that the center of mass determined from the surface mass density map traces the galaxy's gravitational potential. Any clumps located within 0.24 arcsec of the center of mass are therefore considered as part of the bulge structures and are removed from the analyses (see Figure \ref{fig:clumpy_sfg} for examples of these bulge-like components). Finally, we define clumpy galaxies as those with detected clumps that contribute to more than 5\% of the total UV luminosity. Only clumps detected in clumpy galaxies are considered for the remainder of the paper.

\subsubsection{Galaxy subtraction} \label{sec:retrieve_mass}

We measure clump stellar masses directly from the stellar mass surface density maps using fixed aperture photometry. 
Since clumps are generally superimposed within a galactic disk, it is essential to account for this underlying background to avoid overestimating their physical properties. 

We assess the robustness of different background subtraction techniques in Appendix \ref{sec:clump_mass_retrieval}. In particular, we test three different methods, including (1) reconstructing the smooth underlying disk by decomposing the mass map into different wavelet scales and removing the contribution of the two lowest scales (similar to \S \ref{sec:find_clumps} when creating a smooth image of the galaxies), (2) modeling the disk by fitting a S\'ersic profile, and (3) estimating the local background via an annular aperture. In general, we find that all three methods reliably recover masses of artificially injected clumps. In the end, we adopt the annular subtraction method as it relies on simpler assumptions of the underlying disk, compared to parametric modeling approaches such as a S\'ersic profile. Other studies such as \cite{Sattari2023} have also used the annular background subtraction to estimate physical properties of clumps. Overall, performing background subtraction generally leads to correction of ${\sim}0.2 ~\mathrm{dex}$ per clump. 

As the clumps in this study are typically unresolved and will have a FWHM of $~{\sim}0.16^{\prime\prime}$, we use an aperture of radius $0.2^{\prime\prime}$ to extract clump masses. This aperture size encloses the majority of the clump's mass (and similarly, light) while minimizing contamination from the underlying disk. Local background estimation is done by calculating the mean stellar mass density between two annuli at 3.5 and 5 pixels around the clumps. We also mask other clumps in the galaxy image while measuring the local stellar mass density to avoid contamination. The local stellar mass density is subtracted from the aperture measurement to isolate the stellar mass associated with each clump. We account for magnification, if any, by dividing the retrieved clump mass by $\mu$. 

We note that clump masses can be alternatively retrieved by fitting the SED to photometry measured directly from the multi-band images. To compare this approach to extracting the clump masses from the two-dimensional mass maps, we perform similar aperture photometry on images using the same background subtraction procedure. Here, magnification is accounted by dividing each flux by the corresponding $\mu$. We find that clump masses from the two approaches are consistent, with a scatter of ${\sim}0.3~\mathrm{dex}$. 

\begin{figure*}[!t]
    \centering
    \includegraphics[width=\linewidth]{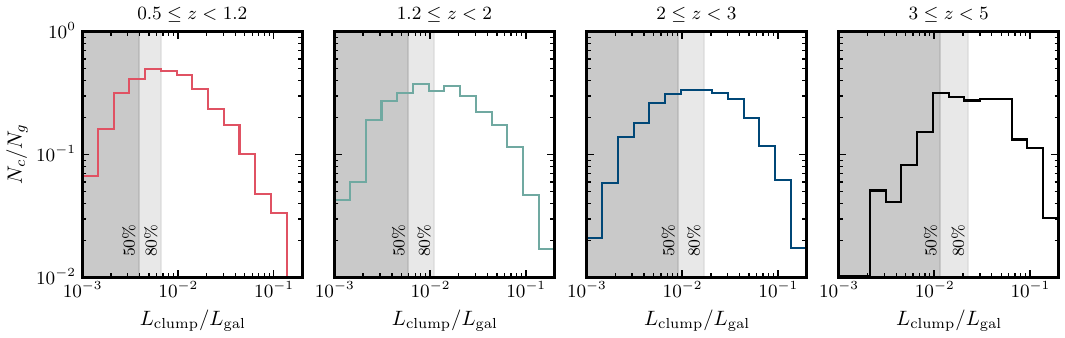}
    \caption{The fractional luminosity function of clumps as a function of redshift. In general, we find that the distribution is shifted toward higher $L_\mathrm{clump}/L_\mathrm{gal}$ values at higher redshifts as fainter clumps become harder to detect. The gray regions denote the 50\% and 80\% completeness limit determined by simulating clumpy galaxies and determining how many clumps are retrieved back.}
    \label{fig:lf}
\end{figure*}

\subsubsection{Clump completeness} \label{sec:completeness}

Understanding our clump detection thresholds is a crucial step towards interpreting our results, including the clump stellar mass function. We determine the clump detection limit by injecting clumps into galaxy images across all available filters and applying our spatially resolved SED-fitting and clump-finding algorithm to quantify our clump recovery.

In Figure \ref{fig:lf}, we first look at the distribution of all clumps detected as a function of their luminosity compared to the host galaxy's luminosity (i.e., the fractional luminosity function). We find that we are able to detect fainter clumps compared to the host galaxy's luminosity at lower redshifts. However, at higher redshifts, the distribution is shifted to higher $L_\mathrm{clump}/L_\mathrm{gal}$ values as clumps become harder to detect. To determine the completeness of our detection, we simulate clumpy galaxies by injecting clumps into them. Rather than inject clumps from an assumed set of parameters, we instead sample directly from a parent sample of detected clumps and inject them into different galaxies. This approach ensures that the simulated clumps reflect the intrinsic range of observable clump properties. Since clumps are detected based on their rest-frame UV brightness, we explore mass detection limits by randomly scaling clump fluxes so that their final mass span between $7.4<\log(M_{c,}/M_\odot)<8.6$. 

The parent sample of clumps used for simulating clumpy galaxies consists of clumps detected between $0.5<z<4$. We limit this sample to clumps with masses at $7.4<\log(M_{c,}/M_\odot)<8.6$ to avoid selecting either clumps that are well below the detection limit, or brighter clumps that could be satellites. To simulate clumpy galaxies, we randomly select galaxies from our galaxy sample. For each simulated galaxy, we inject ``donor" clumps from the parent clump sample whose redshifts are within $\Delta z=1$ of the simulated galaxy's redshift. The redshift constraint here ensures that these clumps are representative of those observed at similar epoch. Since these donor clumps are generally at a different redshift compared to the target galaxy, we shift their spectral energy distributions accordingly to the new redshift. The SED of clumps is obtained from \textsc{Dense Basis}, using a $0.2^{\prime\prime}$-diameter circular aperture centered on the clump, with background subtraction applied in each filter (see Appendix \ref{sec:clump_mass_retrieval}). 
\noindent The flux of the redshifted SED through each JWST filter is measured with \textsc{Dense Basis}, which account for the filter response. These are then injected into the corresponding images of the simulated galaxy. Each clump is modeled as a 2D elliptical Gaussian with semi-minor axis of FWHM of $0.16^{\prime\prime}$, and is scaled so that the total flux is within an aperture of $0.2^{\prime\prime}$. We randomize the ellipticity between $0.1<e<0.2$ and the position angle. For each simulated galaxy, up to four donor clumps are injected into them. The uncertainties of the clump mass completeness are determined by performing twenty independent realizations of this injection-recovery procedure. In each realization, there are approximately 200 simulated galaxies with ${\sim}800$ injected clumps.

\begin{figure*}[!t]
    \centering
    \includegraphics[width=\linewidth]{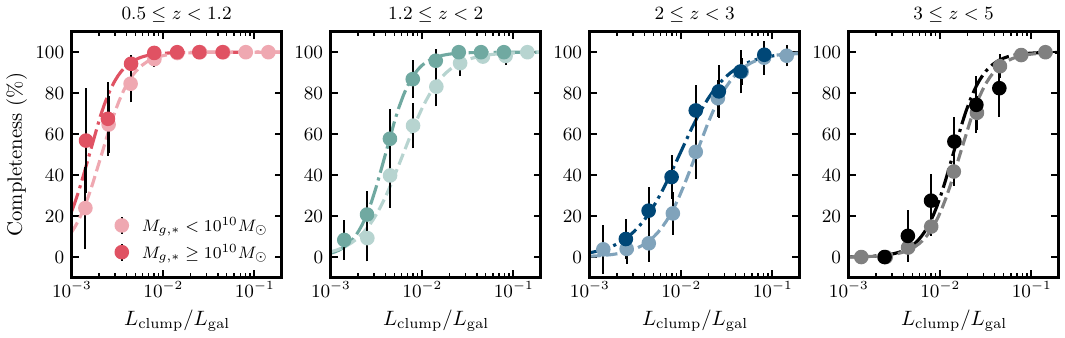}
    \includegraphics[width=\linewidth]{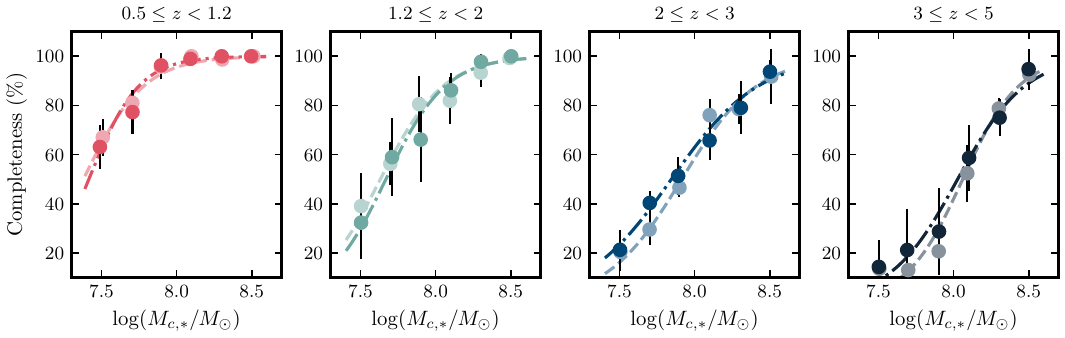}
    \caption{The completeness limit as a function of host galaxy mass and redshift bins. In the top panels, we show the completeness as a function of the fractional luminosity of clumps. The bottom panels show the completeness as a function of clump stellar mass. We determine the completeness by injecting clumps into galaxy images and rerunning our SED fitting and clump detection pipelines (see text for details). In general, we find that the clump completeness is similar for both galaxy mass bins.}
    \label{fig:completeness}
\end{figure*}

The top panels of Figure \ref{fig:completeness} show the completeness as a function of the fractional rest-frame UV luminosity in clumps per redshift bin, separated by galaxy masses. In general, we find that the completeness is lower for less-massive galaxies at fixed fractional clumpy luminosity. This arises because lower mass galaxies will have lower total luminosities, so clumps corresponding to a given fraction of that luminosity are intrinsically fainter and therefore more difficult to detect. Similarly, the bottom panels of Figure \ref{fig:completeness} show the completeness as a function of clump mass. We model the completeness as a sigmoid function, described as, 
\begin{equation} \label{eqn:sigmoid}
    C(M_{c,*}) = \frac{100}{1 + e^{-k(M_{c,*} - M_o)}},
\end{equation}
\noindent where $k$ and $M_o$ are parameters. At clump masses of $\log(M_{c,*}/M_\odot)\sim8.0$, the clump mass completeness is at $80\%$. Given the uncertainties, we find that the clump mass completeness is similar for both low- and high-mass galaxies. However, we note that the mass completeness seems lower for massive galaxies at lower redshifts. This could be driven due to the contrast of the image, where the light from a clump is weaker compared to more massive galaxies.

\section{The Fraction of Clumpy Galaxies} \label{sec:clumpyfrac}

\begin{figure}[t!]
    \centering
    \includegraphics[width=\linewidth]{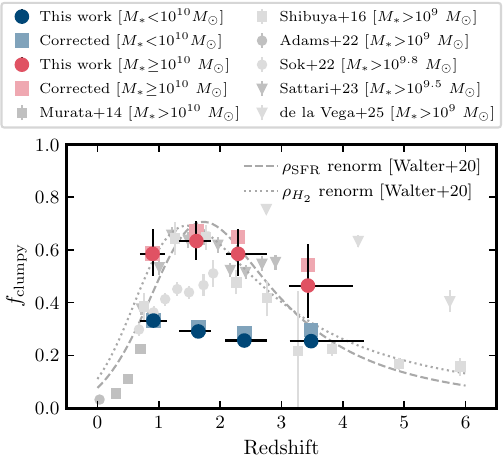}
    \caption{The fraction of clumpy star-forming galaxies for two different galaxy mass bins, where the red markers denote galaxies with $\log(M_{g,*}/M_\odot)>10$ and the blue markers denotes galaxies with $8.5<\log(M_{g,*}/M_\odot)<10$. We further correct the clumpy fraction based on a Monte Carlo approach where we sample clumps from both the corrected and observed fractional luminosity function to determine classification incompleteness (see text for details). We also compile and show previous clumpy fraction measurements, shown as the lighter gray markers. The dashed and dotted line show the cosmic SFR and molecular gas density from \cite{Walter2020}, renormalized to fit our measured clumpy fraction of galaxies with $\log(M_{g,*}/M_\odot)>10$.}
    \label{fig:combined_clumpy_fraction}
\end{figure}

In this section, we examine the abundance of clumpy galaxies between $0.5<z<5$. Clumpy galaxies are defined as those with clumps that contribute to more than 5\% of their total UV luminosity. We denote the $f_\mathrm{clumpy}$ as the fraction of star-forming galaxies that are clumpy. 
Figure \ref{fig:combined_clumpy_fraction} shows the evolution of the fraction of clumpy galaxies at different redshift bins, and the error bars represent the $1\sigma$ uncertainties assuming binomial statistics. We find differences in the redshift evolution based on galaxy masses. Specifically, massive galaxies with $\log(M_{g,*}/M_\odot)\geq10$ exhibit a stronger evolution with redshift relative to less massive galaxies with $8.5<\log(M_{g,*}/M_\odot)<10$. The clumpy fraction for massive galaxies peaks at around 60\% near $z\sim2$ and remains at around 40\% at $z\sim4$. For less massive galaxies, the clumpy fraction is observed to be at 30\% for all redshift bins.

The trend in the redshift evolution of the clumpy fraction can be driven due to incompleteness, where clumps become fainter and harder to detect at higher redshift. To estimate a correction to the clumpy fraction, we determine the incompleteness of our classification based on a Monte Carlo approach. For each redshift bin, we integrate both the observed and corrected fractional luminosity function from $\log(L_\mathrm{clump}/L_\mathrm{gal}) > -2$ (see \S \ref{sec:completeness}) to determine the expected number of clumps per galaxy. In each realization, we then sample that number of clumps directly from the fractional luminosity functions to determine if the sampled clumps add up to more than 5\%. This provides an estimate for the classification completeness $X$ of our sample based on the corrected fractional luminosity functions. The corrected $f_\mathrm{clumpy}$ is then calculated following a similar equation in \cite{Guo2015}, 
\begin{equation}
    f^\mathrm{corr}_\mathrm{clumpy} = f^\mathrm{old}_\mathrm{clumpy} + \big(\frac{1}{X} - 1\big)f^\mathrm{old}_\mathrm{clumpy}.
\end{equation}
\noindent The second term on the right hand side accounts for the fraction of galaxy that may have been clumpy but classified as non-clumpy due to non-detections of clumps. In general, we find that the classification completeness is around 80-95\% depending on the redshift bins. These values are likely lower if clumps of lower fractional luminosity are included, however we limit the correction to clumps with $\log(L_\mathrm{clump}/L_\mathrm{gal}) > -2$, where the completeness is approximately at 50\% and the correction is more reliable.

We also compare our results to other studies. While the differences in the normalization of the clumpy fraction can be affected by choices in sample selection and clump detection thresholding, we generally find that the overall redshift evolution of the clumpy fraction is consistent with previous studies for galaxies of similar mass bins. This is shown as the gray markers in Figure \ref{fig:combined_clumpy_fraction}. In particular, studies of lower-redshift clumpy galaxies suggest a steep decline in the clumpy fraction at $z<1$ \citep{Murata2014, Hinojosa2016, Adams2022}. \cite{Shibuya2016} further noted that the clumpy fraction peaks near cosmic noon and declines at earlier times, although the true clumpy fraction may drop more slowly at $z>2$ due to incompleteness. They suggested that the clumpy fraction broadly traces the evolution of the cosmic star formation rate density (SFRD, \citealt{Madau2014}). This redshift evolution is also observed by \cite{Vega2025}. Their higher clumpy fraction arises from using a lower threshold at $2\%$ compared to the $5\%$ of this study. 

We reproduce the redshift evolution by fitting our uncorrected clumpy fraction of galaxies with $\log(M_{g,*}/M_\odot)\geq10$ using the empirical prescription from \cite{Madau2014}, 
\begin{equation}
    f_\mathrm{clumpy} = a \frac{(1+z)^{b}}{1+[(1+z)/c]^d},
\end{equation}
\noindent where we fix the parameters $b$, $c$ and $d$ to the values reported in \cite{Walter2020}, and only refit for $a$ to renormalize the best-fit curve. This is plotted as the dashed gray line in Figure \ref{fig:combined_clumpy_fraction}. Similarly, the renormalized molecular gas density is shown as a dotted line. The observed clumpy fraction broadly follows the redshift evolution of both the molecular and SFR densities. 

Across all redshifts, we find that massive galaxies are more likely to be clumpy than low mass galaxies. While previous studies (e.g., \citealt{Sok2022, Sattari2023}) have reported a negative correlation between galaxy mass and clumpy fraction, it should be noted that our sample contains galaxies with stellar masses approximately an order of magnitude lower than those previous works. For example, the clumpy fraction as observed in both \cite{Murata2014} and \cite{Huertas2020} is seemingly peaked at $\log(M_*/M_\odot) \sim 10$ before declining at lower masses. In addition to stellar masses, the clumpy fraction is also dependent on additional parameter such as star formation rate (e.g., \citealt{Murata2014, Sok2022}).

\section{Clump Properties} \label{sec:clump_prop}

In this section, we examine the relationship between clump properties and the global properties of their host galaxies, with a particular focus on stellar mass. We further analyze the clump mass distribution and how it evolves with redshift. As noted in \S \ref{sec:find_clumps}, these analyses are restricted to clumps and do not include bulge-like components detected in galaxies.

\subsection{Clumps/Host Galaxies Relations}

\begin{figure}[t!]
    \centering
    \includegraphics[width=\linewidth]{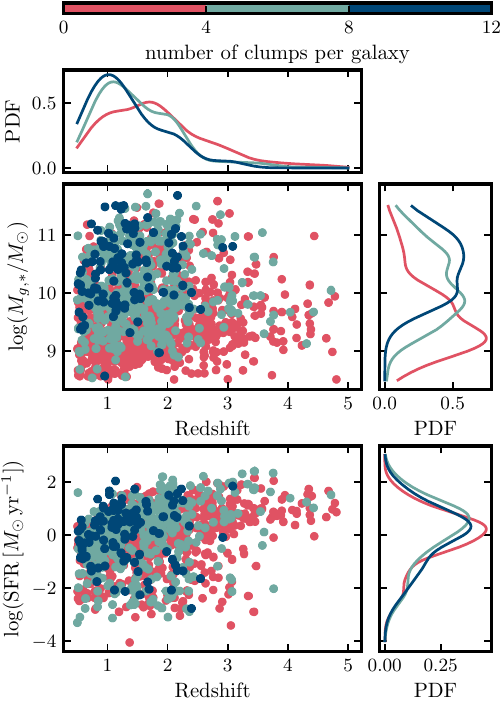}
    \caption{Number of clumps in clumpy galaxies as a function of their host galaxy mass and SFR. The marginal probability distributions, color-coded by the number of clumps, are shown in the top and right panels. Massive galaxies tend to host more clumps relative to less massive galaxies, however we find no preference with the total SFR.}
    \label{fig:clump_counts}
\end{figure}

\subsubsection{Number of clumps per galaxy}

Figure \ref{fig:clump_counts} shows the number of clumps found in galaxies as a function of the host galaxy mass and star formation rate. The marginal probability distributions are also shown the top and the right panels. We find that the number of clumps is generally higher for more massive galaxies at fixed redshift, similar to what was found in \cite{Soto2017}.
When looking at star formation rates, we find that galaxies with higher star formation rate do not seem to preferentially host more clumps compared to those with lower SFRs. This is seen more clearly in the lower-right panel, which shows the marginal probability distribution of SFR, color-coded by the number of clumps per galaxy. In general, the distributions peak at similar SFR values regardless of clump count.
Above $z>3$, the number of detected clumps decreases. However, this is likely driven by detection limits, where clumps of a fixed fractional luminosity of their host galaxies are less complete.  

\begin{figure}[!t]
    \centering
    \includegraphics[width=\linewidth]{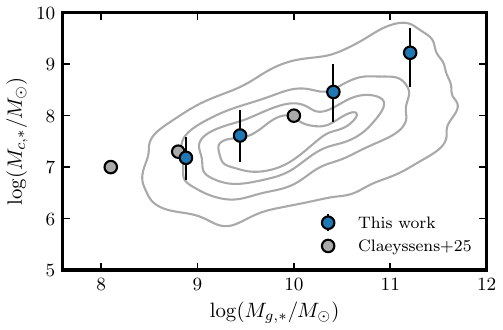}
    \caption{The relationship between clump masses and their host galaxy masses. The contours show the full distribution of the clump sample. To further investigate the relationship between clump mass and host galaxy mass, we select the most massive clump in each galaxy, and compute the median mass of these clumps within bins of host galaxy stellar mass. These median values are shown as blue markers. Similar correlation between the most massive clump mass and host galaxy mass is also observed by \cite{Claeyssens2025}, who performed analyses on a compilation of clumps found in lensed sources between $0.6<z<5$. }
    \label{fig:clump_mass_distribution}
\end{figure}

\subsubsection{Clump masses} \label{sec:mass_relation}

Figure \ref{fig:clump_mass_distribution} shows the distribution between clump mass and the host galaxy mass for all clumps in our sample as the contours. We also examine the relationship between the mass of the most massive clump in each galaxy and its host galaxy mass. This is shown as the blue markers, which represent the median mass of these most massive clumps in host galaxy mass bins of $10^8$, $10^9$, $10^{10}$, $10^{11}$, and $10^{12}$ $M_\odot$. Similar trends have been reported in previous studies; for example, \cite{Claeyssens2025} also found a strong positive correlation between the maximum clump mass and the host galaxy mass (reproduced in Figure \ref{fig:clump_mass_distribution}). We further examine the correlation between the maximum clump mass and host galaxy mass by computing the Pearson correlation coefficient on the binned data points. The Pearson test measures the linear correlation between two variables, with coefficients ranging from -1 to 1. A value of one implies positive correlation. We find a strong and significant corelation, with a Pearson coefficient of 0.73 and p-value less than 0.001. The p-value here indicates that the probability of obtaining such a correlation by chance, if no correlation existed, is less than 0.1\%.

In general, the correlation between clump mass and host galaxy mass, as shown by the contours of Figure \ref{fig:clump_mass_distribution}, suggests a fundamental link between clump formation and the broader physical context of their host galaxy. In particular, while massive galaxies tend to host the most massive clumps, the contours also suggest that low-mass clumps are are less common in massive galaxies.
This becomes clearer in \S \ref{sec:csmf} when we examine the clump mass distribution as a function of galaxy mass and find that galaxies with $\log(M_{g,*}/M_\odot)>10$ appear to have a flatter slope relative to less massive galaxies. This suggests that massive galaxies have a higher fraction of massive clumps relative to lower mass clumps.  
We do not expect the absence of low-mass clumps in massive galaxies to be primarily driven by detection biases as we note that our completeness limit is generally similar for all redshift bins at the stellar mass range used for our analysis (see Figure \ref{fig:completeness}). The next section shows that the clump mass distribution across galaxy masses are still different even with a completeness correction. 

\subsection{The Clump Stellar Mass Function} \label{sec:csmf}

The clump stellar mass function (cSMF) contains crucial insights on the formation mechanism of clumps. Observations in the local Universe indicate that stellar mass distributions of star clusters often follow a power-law form with a slope of $\alpha = -2$ \citep[e.g.,][]{Krumholz2019, Adamo2020}, which is expected for structures formed via hierarchical, scale-free fragmentation (e.g., \citealt{Hopkins2013, Guszejnov2018}).

Due to the limited number of clumps identified in individual galaxies, a common practice is to combine clumps across multiple galaxies to analyze their aggregated clump stellar mass function. In the following sections, we investigate the redshift evolution of the aggregated cSMF, as well as the cSMFs for two different galaxy mass bins. 
In order to measure the clump stellar mass function as a function of redshift, we define the fiducial redshift bin edges at 0.5, 1.2, 2, 3 and 5. We also account for incompleteness in the following analyses. Our sample reaches an 80\% completeness limit for $\log(M_{c,*}/M_\odot) \sim 7.7-8.3$ between our lowest and highest bins. 

\subsubsection{Fitting the cSMF}
The clump mass function is modeled as a power-law, 
\begin{equation}
    \frac{dN}{dM_{c,*}} \propto M_{c,*}^{\alpha}.
\end{equation}
The distribution is first constructed by using uniform bin width of 0.3 dex spanning $7.5<\log(M_{c,*}/M_\odot)<9.6$. Within each bin, we count the number of clumps and measure the mean clump mass. We apply the completeness correction and restrict the fit to the bins where the average clump mass lies above the 50\% completeness limit. The uncertainties of each bin are determined by bootstrap resampling. Specifically, we generate 1000 realizations of the clump sample by randomly sampling the clump masses with replacement. We then compute the binned cSMF for each realization and adopt the dispersion of the resulting distributions as the $1\sigma$ uncertainty.

We also fit the clump stellar mass function using a Bayesian approach. We explore the posterior distribution of $\alpha$ using a Markov Chain Monte Carlo (MCMC) method implemented in \textsc{emcee} \citep{Foreman2013}. The log-likelihood function follows a similar form to the one outlined in \cite{Giunchi2025}, 

\begin{equation}
\ln \mathcal{L}(\alpha) = \sum_{i=1}^{N_\mathrm{cl}}
\ln \left[
\frac{\int \mathrm{MF}(x;\alpha)\,C(x)\,G_i(y_i \mid x)\,dx}
{\int \mathrm{MF}(x;\alpha)\,C(x)\,dx}
\right],
\end{equation}
where $N_\mathrm{{cl}}$ is the number of clumps above the 50\% mass completeness, $C(x)$ is the completeness (see \S\ref{sec:completeness}), $x=\log(M_{c,*})$, $G_i(y_i \mid x)$ is the measurement uncertainties of the $i$-th clump mass described as a Gaussian, and $\mathrm{MF}$ is the clump mass function. Similarly, we limit the analyses to clump masses that are above our 50\% completeness limit. The values for both methods are provided in Table \ref{tab:clump_all_slopes}. Both methods yield similar results that are within uncertainties. All quoted cSMF slopes in the paper are based on the binned method.

\subsubsection{cSMF for different galaxy mass bins}


\begin{figure*}
    \centering
    \includegraphics[width=\linewidth]{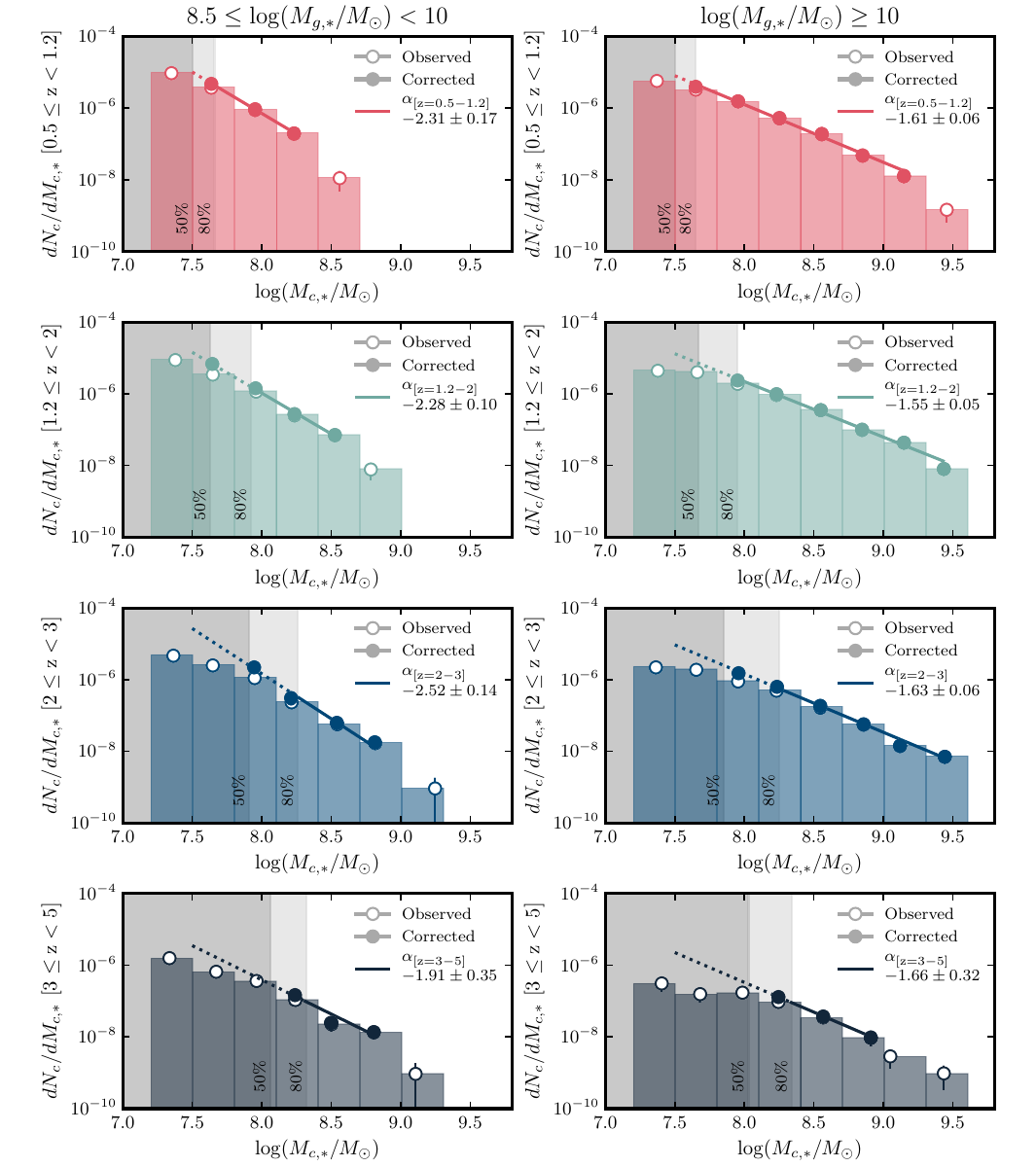}
    \caption{The clump stellar mass function (cSMF) in relation to redshifts and galaxy masses. Open markers denote the observed cSMF, while solid markers show the corrected cSMF, which has a correction factor based on our clump mass completeness limit. The darker and lighter gray regions denote where we are mass complete at 50\% and 80\%, respectively. The slope of the mass function is only fitted using clump mass bins above our 50\% completeness limit and bins that contain more than five clumps.}
    \label{fig:csmf_sep}
\end{figure*}

In this section, we divide the clump mass function by two host galaxy masses, using the fiducial mass bin of $8.5<\log(M_{g,*}/M_\odot)<10$ and $\log(M_{g,*}/M_\odot)\geq10$ to represent the low- and high-galaxy mass bins, respectively. Figure \ref{fig:csmf_sep} presents the resulting cSMFs across different redshift and galaxy mass bins. The histograms show the observed (uncorrected) cSMFs, while the solid markers denote the mass-corrected distributions. 


We find that the power-law slope of the cSMF is generally flatter for more massive galaxies compared to less massive galaxies, with typical values of $\alpha_\mathrm{[high-mass]} \sim -1.6$ compared to $\alpha_\mathrm{[low-mass]} \sim -2.3$.
This result is consistent with \S \ref{sec:mass_relation}, where we find that massive galaxies tends to host more massive clumps, likely contributing to the flatter cSMF slopes. We also note that the steeper slopes found in less massive galaxies could be partly driven by the physical limit that clump mass cannot be more massive than their host galaxy mass.
The cSMF slopes also show little to no evolution with redshift for both galaxy bins, given the large uncertainty associated with each measurement.

We also fit the clump stellar mass functions using only clump mass bins above the 80\% completeness limit to investigate whether the observed differences in slope are driven by the incompleteness correction. This is summarized in Table \ref{tab:clump_all_slopes}. Some redshift bins, particularly for the low galaxy mass bin, have a limited number of clumps, so in these cases the slope is constrained by only two data points. Noting this caveat, we generally find that the results are broadly consistent to the mass completeness correction to the 50\% threshold. That is, we find that massive galaxies tend to have flatter slope compared to lower mass systems, with little to no redshift evolution. 


\subsubsection{Combined cSMF}

\begin{deluxetable*}{lccccc}[t!]
\tablecaption{The slope of the clump stellar mass function as a function of redshift and galaxy mass. We include fitted values using both a binned and Bayesian approach. We account for incompleteness down to the 50\% limit. In the binned method, uncorrected slopes are measured using only data points above the 80\% limit, with no correction applied. \label{tab:clump_all_slopes}}
\tablehead{
\colhead{Galaxy Mass Bin} &
\colhead{Method} &
\colhead{$0.5<z<1.2$} &
\colhead{$1.2<z<2$} &
\colhead{$2<z<3$} &
\colhead{$3<z<5$}
}
\startdata
$8.5 \leq \log(M_{g,*}/M_\odot) < 10$  & Binned + Uncorrected  & $-2.47\pm0.37$&$-2.18\pm0.19$&$-1.91\pm0.65$&$-0.77\pm0.77$ \\
          & Binned + Corrected   & $-2.36\pm0.17$ & $-2.27\pm0.10$ & $-2.52\pm0.14$ &$-1.89\pm0.33$ \\
          & Bayesian  & $-2.47\pm0.10$&$-2.22\pm0.12$&$-2.52\pm0.19$&$-2.36\pm0.46$ \\
$\log(M_{g,*}/M_\odot) > 10$ &  Binned + Uncorrected & $-1.66\pm0.08$&$-1.58\pm0.06$&$-1.57\pm0.13$&$-1.62\pm0.67$ \\
          & Binned + Corrected & $-1.61\pm0.06$&$-1.56\pm0.05$&$-1.61\pm0.06$&$-1.66\pm0.33$  \\
          & Bayesian  & $-1.59\pm0.05$&$-1.56\pm0.04$&$-1.60\pm0.07$&$-1.71\pm0.20$ \\
Combined  & Binned + Uncorrected & $-1.93\pm0.12$&$-1.69\pm0.06$&$-1.67\pm0.08$&$-1.57\pm0.24$ \\
        & Binned + Corrected & $-1.87\pm0.05$&$-1.70\pm0.05$&$-1.88\pm0.05$&$-1.81\pm0.24$ \\
          & Bayesian  & $-1.91\pm0.05$&$-1.73\pm0.04$&$-1.87\pm0.06$&$-2.04\pm0.16$ \\
\enddata
\end{deluxetable*}

\begin{figure}
    \centering
    \includegraphics[width=\linewidth]{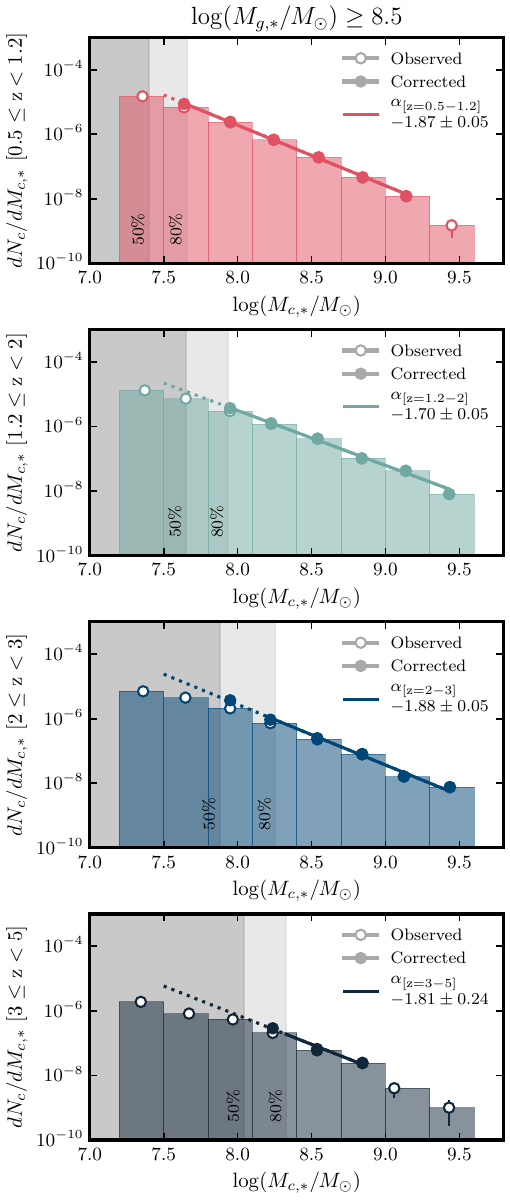}
    \caption{Similar to Fig. \ref{fig:csmf_sep} but for the combined cSMF. Before combining, a correction factor is applied to each cSMF to account for any completeness differences between the galaxy mass bins. The darker and lighter gray region denote where the mass completeness is 50\% and 80\% complete, respective.  }
    \label{fig:csmf}
\end{figure}

Figure \ref{fig:csmf} shows the combined cSMF for the different redshift bins. Again, the sloped line shows the fit to the cSMF, where the solid portion of the line highlights the range of masses where we are mass-complete. In general, we find power-law slopes that are slighter lower than $\alpha = -2$, but is consistent to what are reported by previous studies. Specifically, the typical reported slope ranges between $-1.5$ to $-2$ in other studies at redshift of 2 (e.g., \citealt{Dessauges2018, Kalita2025b_nir_clumps, Kalita2025a_csmf}). \cite{Ambachew2022} found a flatter slope of $-1.4$ for clumps observed in the DYNAMO sample. This sample consists of galaxies with rotating, marginally stable disk at $z\sim0.1$, which are proposed to be analogs of high-$z$ clumpy galaxies (e.g., \citealt{Fisher2017}). Using the VELA simulations, \cite{Huertas2020} reported a shallower slope of $\alpha = -1.5$. However, they found an even shallower slope of $\lesssim-1$ for clumps in galaxies with $\log(M_{g,*}/M_\odot) > 9$ at $1<z<3$ in CANDELS.

In general, we find no redshift evolution in the slope. This differs from our earlier results where we find a slight redshift evolution that gets flatter with redshift for galaxies of lower mass. To assess the impact of completeness corrections on the result, we also fit the slope using clump mass bins that are above the 80\% mass completeness limit (e.g., $\log(M_{g,*}/M_\odot)\gtrsim8$). Without any correction, we find similar slope values. There is a more noticeable redshift evolution in the uncorrected cSMF slopes, where it flattens strongly toward higher redshifts. This is likely driven by detection limits, where less massive clumps are becoming increasingly difficult to detect at higher $z$.

\begin{figure}
    \centering
    \includegraphics[width=\linewidth]{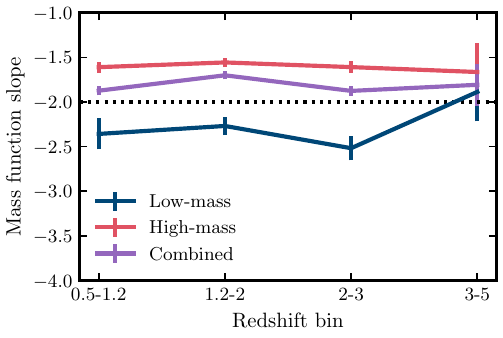}
    \caption{The redshift evolution of the cSMF based on different galaxy mass bins.}
    \label{fig:csmf_evolution}
\end{figure}

A summary of the slopes is provided in Table~\ref{tab:clump_all_slopes}, and Figure \ref{fig:csmf_evolution} shows the redshift evolution of the slopes for the different galaxy mass bins. When aggregating the entire clump sample, we find that the slope is fairly consistent throughout redshift with a slope of $\alpha\sim-2$. Massive galaxies tend to have flatter slopes compared to less massive galaxies, with $\alpha_\mathrm{[high-mass]} \sim -1.6$ compared to $\alpha_\mathrm{[low-mass]} \sim -2.3$.  This could reflect a mixing of galaxy population. For example, while differing slope evolution in low- and high-mass galaxies are observed, combining the relative abundances of clumps in different populations can average out these underlying trends. In particular, the cSMF slopes of the combined sample are closer to the high galaxy mass bin, where there are more detected clumps. 

\section{Discussion}

In this work, we identify clumps in star-forming galaxies in CANUCS between $0.5<z<5$. We measure the slope of the cSMF, limiting our analyses to clump masses above the 50\% completeness limit and applying corrections for incompleteness. When aggregating the clump sample for all galaxies, our results show little evolution in the slope of the clump mass function between $0.5<z<5$. Our finding is consistent with what is reported by \cite{Dessauges2018, Claeyssens2025}. In the latter case, \cite{Claeyssens2025} compiled one of the largest sample of ${\sim}1700$ lensed clumps and found that their cumulative mass functions are largely consistent with a power-law slope of $\alpha=-2$ for $0.7<z<5.5$. As discussed in \cite{Dessauges2018}, these results are consistent with a scenario where clumps form predominantly \textit{in-situ}, with their mass distribution arising from turbulence-induced gravitational fragmentation. The comparable cSMF slopes across redshifts suggest similar formation conditions.


However, our findings suggest differences in the cSMF for the two different galaxy mass bins, where galaxies with masses of $\log(M_{g,*}/M_\odot) > 10$ have slopes of $\alpha_\mathrm{[high-mass]} \sim -1.6$ while galaxies with masses of $8.5 < \log(M_{g,*}/M_\odot) < 10$ have steeper slopes of $\alpha_\mathrm{[low-mass]} \sim -2.3$ . We find that there is little redshift evolution in the slopes for both mass bins. Similarly, \cite{Huertas2020} found flatter slope in massive galaxies in the CANDELS sample, but attributed this to incompleteness with low-mass clumps being less complete in massive galaxies. However, we find that this difference cannot be fully explained even with a completeness correction in our analyses. 

Taken together, the observed trends of the cSMF likely reflect an interplay between the conditions for clump formation and the mechanisms that destroy them over cosmic time. A flatter slope for a cSMF could reflect either an intrinsically more top-heavy distribution of clumps at formation, or a universal initial cSMF with preferential destruction of low-mass clumps over time, or a combination of both. We explore some of these possibilities in the following discussion. First, we consider potential mechanisms for clump formation and how it may flatten the cSMF slopes, focusing on the roles of gas and mergers. We then examine how clump destruction can also flatten the cSMFs by preferentially disrupting clumps of lower masses.

\subsection{Clump formation origins}
\subsubsection{Are the slopes of the cSMF driven by gas fraction?}
The prevalent theory for clump formation is through violent disk instability, regulated by the Toomre \textit{Q} parameter, which depends on the dynamical state of the disk and its gas content (e.g., \citealt{Dekel2009_clumpy_disks, Ceverino2010}). Consequently, gas fraction plays a role in setting the conditions for instabilities and the resulting clump population. For example, several simulations suggest that higher gas fractions flatten the cSMF slope by enabling the formation of more massive clumps \citep[e.g.,][]{Fensch2021, Renaud2024}.

\begin{figure}[!t]
    \centering
    \includegraphics[width=\linewidth]{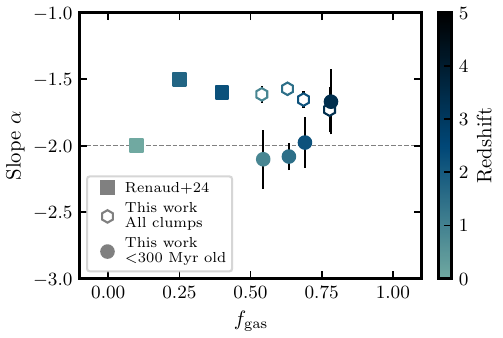}
    \caption{The relationship between the cSMF slope and gas fraction. The data from \cite{Renaud2024} represents simulations of isolated galaxies at varying $f_\mathrm{gas}$, and are analogous of the Milky-Way at different redshifts, with mass of ${\sim}10^{10.7}$ solar masses. In comparison, we plot the slope of cSMF for galaxies between $\log(M_{g,*}/M_\odot)\geq10$ for the different redshift bins and their corresponding $f_\mathrm{gas}$, derived using the relation from \cite{Wang2022}. The open circle markers denote the cSMF slopes derived using all clumps, while the filled circle markers only account for clumps that are ${<}300$ Myr old. The filled circle markers have been shifted by 0.005 in $f_\mathrm{gas}$ for visual clarity.}
    \label{fig:slope_vs_fgas}
\end{figure}

We further explore this relation using our observed slopes. We first limit the discussion to galaxies with stellar masses above $10^{10}$ solar masses due to the larger number of detected clumps in this mass bin. While we do not have a direct measurement of molecular gas, we instead estimate the gas fraction based on the mean stellar masses and SFRs for each of our redshift bin, following the relation described in \cite{Wang2022}. Figure \ref{fig:slope_vs_fgas} therefore shows the relationship between increasing gas fraction and the cSMF slope, as well as values reported in \cite{Renaud2024}. At fixed masses, the galaxies in our redshift bins generally have a higher gas fraction with increasing redshifts. Contrast to what is reported by \cite{Renaud2024}, we find little correlation between gas fraction and the cSMF slope (open markers). 

However, \cite{Fensch2021} noted the cSMF slopes tend to flatten from an initial cSMF as clumps evolve and gain mass in their simulation. It could therefore be that initial cSMF is dependent on gas fraction, where more massive clumps are formed in gas-rich disks at beginning, but population mixing as clumps evolve will average out the cSMF slope. To remove the effects of different clump population, we further focus on clumps with ages less than 300 Myr old. We remind the reader that the age is defined to be the lookback time at which 75\% of the total stellar mass had formed, derived from SED fitting with a non-parametric SFH and typically more than ten photometric bands. The age criterion therefore provides a closer probe of the initial clump population while retaining a large clump sample for the analyses. The filled circle markers in Figure \ref{fig:slope_vs_fgas} show the cSMF slope for clumps with ages less than 300 Myr old. We find that the slope of the cSMF of younger clumps appears to correlate with gas fraction, albeit with larger uncertainties.

Contradictory, the clump stellar mass function slopes for low-mass galaxies are steeper than those in high-mass systems, despite potentially having higher molecular gas fractions (i.e., if the relation from \cite{Wang2022} extends to lower masses). 
However, this paradox can be resolved by two effects that regulate clump formation. Firstly, massive clumps are preferentially formed in massive disks that have larger total gas reservoirs. From Toomre analyses, the fragmentation scale in a differentially rotating disk is given by $\lambda_c \sim G\Sigma/\Omega^2$ \citep[e.g.,][Chapter 6.2.3]{Binney2008}, where $G$ is the gravitational constant, $\Sigma$ is the gas density, and $\Omega$ is the angular circular velocity. \cite{Dekel2009_clumpy_disks} showed that this sets the maximum clump scale, with $\lambda_c \equiv R_c \propto \delta R_d$, and the mass of clumps as $M_c \propto \delta^2 M_d$. Here, $\delta$ is the disk-to-total mass fraction, $R_d$ and $M_d$ are the radius and mass of the gas disk, respectively. This scaling implies that more massive disks preferentially host more massive clumps. Secondly, the number of clump formed is related to the dynamical states of the disk, with more clumps typically produced in rotationally supported systems (e.g., \citealt{Orr2024}). Less massive galaxies, which are generally more dispersion-dominated (e.g., \citealt{Reyes2023}), will generally form both fewer and lower mass clumps.

It is important to note a few caveats in this discussion. The simulated galaxies in \cite{Renaud2024} represent Milk-Way analogues, spanning from local universe to $z\sim2.5$, and are isolated systems without any evolutionary, cosmological contexts. Their gas fraction $f_\mathrm{gas}$ also span between $0.1-0.4$. Our galaxies span between $0.5<z<5$, and are estimated to be more gas-rich based on the relation from \cite{Wang2022}, with $0.5<f_\mathrm{gas}<0.7$. While our galaxies are different compared to those in \cite{Renaud2024}, we emphasize that this comparison only aims to investigate how the slope of the cSMF changes in different gas environment, but for galaxies of comparable masses.
While the result suggests a trend between increasing gas fractions and flatter cSMF slope for massive galaxies, this correlation cannot account for the differences in the slope across galaxy mass bins. This points to other factors that affect the shape of the clump mass distribution. 


\subsubsection{Can mergers affect the cSMF?}
In this section, we now consider how mergers, specifically clumps that are of \textit{ex-situ} origins, affect the slope of the cSMF. Clumps of \textit{ex-situ} origins, possibly accreting satellites, generally have higher masses relative to those formed \textit{in-situ} (e.g., \citealt{Mandelker2014, Mandelker2017}), possibly flattening the slope of the mass function. Observational studies, including \cite{Ribeiro2017, Zanella2019}, have also suggested that some clumps are remnants of accreting satellites, due to their extended sizes and larger masses relative to other clump populations. 

The major merger pair fraction is observed to increase from $z=0$ to $z=3$, but remains flat approximately after $z\sim3$ \citep[e.g.,][]{Duan2025, Puskas2025}. If \textit{ex-situ} clumps are typically larger and more massive compared to \textit{in-situ} clumps, this suggests that the cSMF slopes may flatten with redshift when mergers are more frequent.
A visual inspection of Figure \ref{fig:clumpy_sfg} shows a lack of disky galaxy at higher redshifts, where, instead, clumpy galaxies resemble galaxy pairs or satellites. This hints at the possibility a part of the clump populations is of \textit{ex-situ} origins. Indeed, a recent resolved spectroscopic observation with JWST/NIRISS indicates metal-poor gas accretion in cosmic noon galaxies, in which a non-negligible fraction of clumps is observed having lower gas-phase metallicity compared to the host disks \citep{Estrada2025}. Simulations have shown that proto-clumps can form along gas filaments accreting onto the host galaxies \citep{Donkelaar2023}. In these scenarios, such metallicity offsets in clumps point to other formation channels such as accreting gas clumps. 

In addition to gas fraction, clumps of \textit{ex-situ} origins may also contribute to the clump mass distribution by flattening the distribution if they represent a higher proportion of the clump population relative to those formed \textit{in-situ}. Identifying satellites based on imaging alone is challenging, and future works involving large IFU surveys are needed to be able to distinguish \textit{in-situ} and \textit{ex-situ} clumps. 

\begin{figure*}[!t]
    \centering
    \includegraphics[width=\linewidth]{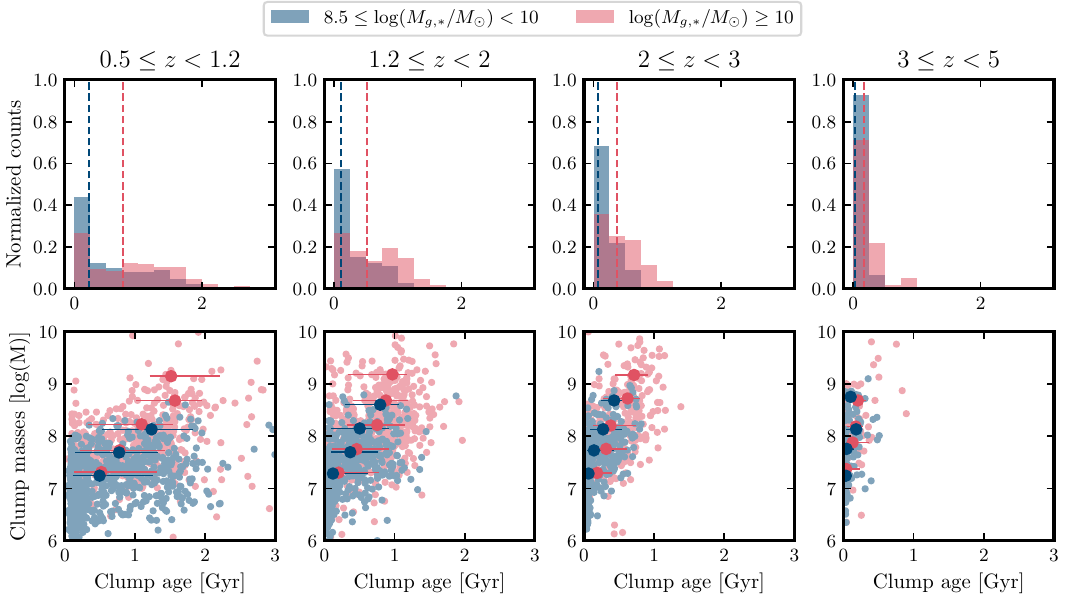}
    \caption{The top panel shows the normalized distribution of clump ages, while the bottom panels show the distribution of clump ages with respect to the their masses. The error bars denote the 16th, 50th and 84th percentile. We find that the age of clumps is correlated with their mass.}
    \label{fig:clump_age}
\end{figure*}

\subsection{Clump destruction} \label{sec:clump_des}
In the previous discussions, we investigate how the slopes of the cSMF are affected by formation mechanisms. Here, we consider whether the cSMF is shaped by processes that destroy clumps. For example, massive galaxies tend to exhibit a flatter clump mass function slope compared to low-mass galaxies, which could result from either the creation of more massive clumps, or from the preferential destruction of low-mass clumps. Feedback processes, such as supernovae and radiation pressure, limit clump survivability, particularly at lower clump masses \citep[e.g.,][]{Moody2014, Mandelker2017, Ginzburg2021, Dekel2022}. Their survivability is also further dependent on clump and disk kinematics, along with bulge and disk mass fraction \citep{Dekel2023}. 

\begin{figure}
    \centering
    \includegraphics[width=\linewidth]{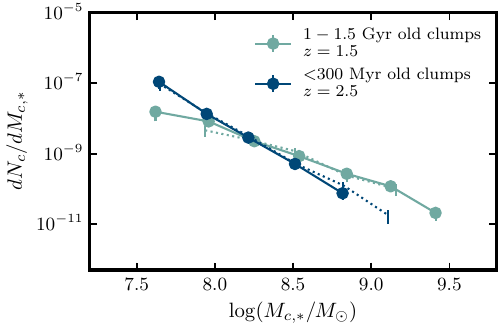}
    \caption{The evolution of the normalized clump mass function for young clumps (${<}300$ Myr old) over $\sim1.5$ Gyr of cosmic time. As clumps evolve, the mass function shows a relative absence of lower mass clumps and the abundance of massive clumps, consistent with a scenario for the preferential destruction of lower mass clumps. The arrow denotes the average mass that would be gained from star formation assuming that the clump’s star formation rate remains constant over this period. This result is independent of the choice for the age selection (e.g., shifting the selection window by 200 Myr does not affect the overall flattening of the clump mass function, which is shown as the dotted line).}
    \label{fig:clump_evolution}
\end{figure}

Here, we test whether lower mass clumps are preferentially disrupted by comparing clump mass and age, using clump ages to define two samples of young and old clumps. The top of Figure \ref{fig:clump_age} shows the distribution of clump ages in low- and high-mass galaxies. The bottom panel shows the relationship between clump mass, their age, and the host galaxy mass. As suggested in the top panel, in addition to the formation of new young clumps, there is also a population of clumps that appear to be aging, as indicated by the shift of the age distribution toward older values at lower redshifts. 

If clump disruption is not mass-preferential, the shape of the cSMF should remain similar between young and old populations. This scenario is tested by selecting a population of young clumps ($<300$ Myr) in our high-mass galaxy bin at $z\sim2.5$ and compare its cSMF to a population of older clumps ($>1$ Gyr and $<1.5$ Gyr) at $z\sim1.5$ at the same galaxy mass bin. Again, we focus on the high galaxy mass bin due to the larger clump sample size. The choice of redshift intervals here is motivated by the fact that there is approximately 1.5 Gyr of cosmic time between $z\sim2.5$ to $z\sim1.5$, roughly tracing the age evolution of the clump populations under the assumption that the young clumps at $z\sim2.5$ will evolve to become those found at $z\sim1.5$. Figure \ref{fig:clump_evolution} shows the cSMF of the young and old clump populations, normalized by the number of clumps in each sample. We find that the mass function slope is steeper for younger clumps, pointing to a higher abundance of low-mass clumps. However, the cSMF becomes flatter for as the clump population ages. This trend supports the scenario in which lower-mass clumps are preferentially disrupted, while massive clumps are more likely to survive and age as their host galaxies evolve. This finding suggests that the clump mass function is not only affect by formation mechanisms, but also by mass-dependent disruption processes. 



\section{Conclusion}

In this work, we use imaging from the CANUCS fields to measure the properties of clumpy star-forming galaxies and their clumps at $0.5<z<5$. Particularly, we measure the evolution of the fraction of clumpy SFGs with cosmic time, and investigate various relationships between clump properties and their host galaxies. We also examine the evolution of the clump mass distribution as a function of galaxy masses and redshifts. We summarize our main finding below:

\begin{itemize}
    \item The fraction of clumpy galaxies ($f_\mathrm{clumpy}$) evolves differently for different galaxy mass bins. For galaxies with $\log(M_{g,*}/M_\odot)\geq10$, the fraction of clumpy SFGs peaks near cosmic noon, and decreases at earlier and later times. We find little to no redshift evolution in the fraction of clumpy SFGs of $8.5\leq\log(M_{g,*}/M_\odot)<10$.
    \item Massive galaxies tend to host more clumps compared to less massive galaxies. However, we do not find the number of clumps per galaxies to be dependent on the integrated star formation rate of the host galaxies.
    \item We find that the mass of clumps appears to correlate with the masses of the host galaxies, implying that massive clumps are generally found in galaxies with massive disks.
    \item When aggregating all clumps, we find little redshift evolution in the clump stellar mass function slope. 
    \item When separating the clump stellar mass function based on galaxy mass bin, we find that galaxies with $\log(M_{g,*}/M_\odot)\geq10$ tend to have flatter slopes ($\alpha_\mathrm{[high-mass]} \sim -1.6$) relative to less massive galaxies ($\alpha_\mathrm{[low-mass]} \sim -2.3$). This suggests either an abundance of massive clumps or an absence of less massive clumps in massive galaxies. 
    \item We also explore the relationship between the cSMF slope and inferred gas fraction. We find that when considering all clumps, the cSMF slope of galaxies with $\log(M_{g,*}/M_\odot)\geq10$ has is uncorrelated to gas fraction. However, when considering younger clumps with age less than 300 Myr old, the slope of the cSMF is flatter with increasing gas fraction. This suggests that higher gas fraction leads to more massive clumps at the time of formation. 
    \item By comparing the slope of a young clump population ($<$300 Myr old) at $z\sim2.5$ to that of an older clump population with ages between 1 to 1.5 Gyr old at $z\sim1.5$, we find that the slope is steeper for the younger clump population. Assuming that the young clump population evolves to become the older clump population, the flattening of the cSMF slope suggests that lower mass clumps are preferentially destroyed, while more massive clumps are likely to survive disruptive feedback processes.
\end{itemize}

Overall, our results show that the aggregated clump mass function remains relatively constant with redshift. However, when considering the cSMF by host galaxy mass bin, we find that it is sensitive to galaxy masses. This likely reflects an interplay between the conditions for clump formation and feedback regulation that disrupt them. 
Our results are consistent with the interpretation that clump formation in massive galaxies ($\log(M_{g,*}/M_\odot) \geq 10$) happens as a result of instabilities and fragmentation within gas-rich disks.
Higher gas fraction can lead to the formation of more massive clumps, which flattens the slope of the clump mass function. However, alternative formation mechanisms, such as accreting satellites, can also be responsible for creating flatter slopes as clumps of \textit{ex-situ} origins are generally considered to be more massive than formed \textit{in-situ}. Together with feedback mechanisms that preferentially disrupt lower-mass clumps, these processes shape the clump mass distribution across different galaxy populations. Additional observational studies will be critical to better constrain the origin of clumps. Particularly, resolved IFU spectroscopy can disentangle the relative contributions of clumps formed via \textit{in-situ} fragmentation and \textit{ex-situ} accretion by exploring the resolved kinematics of clumps with respect to the host disks. Such studies will provide essential insights into the origin of clumps and their role in galaxy evolution. 


\begin{acknowledgments}

This research was enabled by grant 18JWST-GTO1 from the Canadian Space Agency, and Discovery Grant funding from the Natural Sciences and Engineering Research Council (NSERC) of Canada.

MB acknowledges support from the ERC Grant FIRSTLIGHT, Slovenian national research agency ARIS through grants N1-0238 and P1-0188, and the program HST-GO-16667, provided through a grant from the STScI under NASA contract NAS5-26555.

This research used the Canadian Advanced Network For Astronomy Research (CANFAR) operated in partnership with the Canadian Astronomy Data Centre and The Digital Research Alliance of Canada, with support from the National Research Council of Canada, the Canadian Space Agency, CANARIE, and the Canadian Foundation for Innovation. 
\end{acknowledgments}

\appendix

\section{Mass Retrieval with different background subtraction} \label{sec:clump_mass_retrieval}

\begin{figure}[!t]
    \centering
    \includegraphics[width=\linewidth]{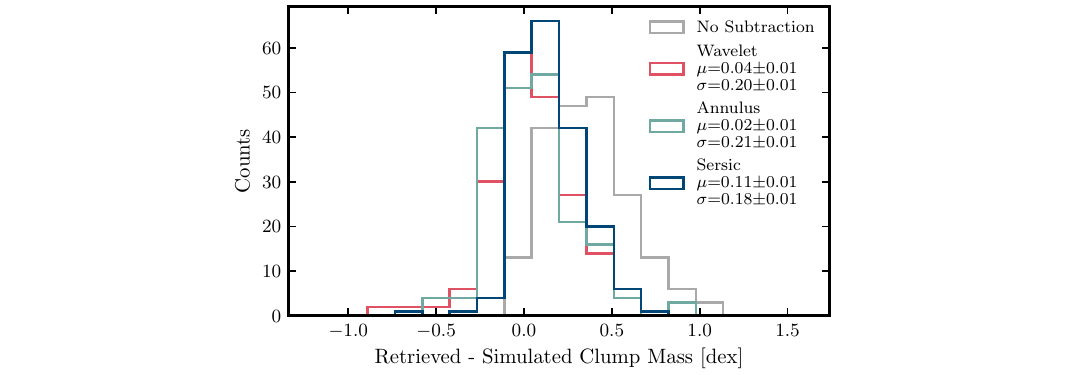}
    \caption{A comparison of different background subtraction methods for retrieving clump masses. In particular, we tested three different methods to estimate the underlying stellar disk; by fitting the disk as a 2D S\'{e}rsic profile, using wavelet transformation to construct a smooth disk (see text for details), and by performing an annular subtraction to estimate local background. We find that all three methods can recover clump masses, but the annular subtraction method generally results with more consistent clump masses. }
    \label{fig:mass_retrieval}
\end{figure}

We test different disk subtraction methods to determine the most reliable approach to extract clump masses. Specifically, we estimate the underlying stellar disk contribution using three techniques: (1) fitting a 2D S\'{e}rsic profile, (2) modeling a smooth disk with wavelet transformation and by setting lower scales to zero, and (3) performing local background subtraction by estimating the mean background within an annular aperture around each clump. 

To model the disk using a S\'{e}rsic profile, we follow the methodology described in \cite{Tan2022}. In brief, we normalize the stellar mass maps to the F444W images, and inject Gaussian noise sampled from the pixel-to-pixel noise of the F444W image. We then use \textsc{statmorph} \citep{Rodriguez2019} to determine the best S\'{e}rsic model, which is then scaled back to the original stellar mass range. For wavelet transformation, similar to what is described in \S \ref{sec:find_clumps}, we decompose the mass maps into 6 different wavelet scales using \textsc{scarlet} and estimate the underlying disk by setting the first two scales to zero. This effectively remove structures with scales less than $0.2^{\prime\prime}$. These estimates of the stellar disk are then subtracted from the stellar mass maps to produce background-subtracted mass maps. Finally, the local background using the annular method involves measuring the mean background within two annuli with radii of 3.5 and 4.5 pixels (corresponding to 0.14 and 0.18 arcsec, respectively). This estimated background is then subtracted from the aperture measurement. 

To test the robustness of each method, we inject simulated clump of mass between $7.0< \log(M_*/M_\odot)<7.8$ into the stellar mass map of nonclumpy galaxies. Nonclumpy galaxies have smoother mass distributions, and therefore present baseline for these tests. We add two clumps to each galaxy at random location, with FWHM of $0.2^{\prime\prime}$. We randomize the position angle of each clumps, with ellipticity ranging between $0.1<e<0.2$. The mass of each injected clump is recovered by measuring the mass within a aperture of radius $0.18^{\prime\prime}$. It should be noted that the underlying stellar disk can be overestimated for some clumps, leading to over-subtraction of the clump mass. This occurs more frequently for both the S\'{e}rsic and wavelet methods.

Figure \ref{fig:mass_retrieval} shows the distribution of the retrieved clump masses relative to their clump mass. In general, all three methods are able to account for the local stellar disk, resulting in typical mass correction of $0.4$ dex. While each method has its strengths and weaknesses, we adopt the annular subtraction method for this work. This is because using wavelets tends to overestimate the disk contribution, while the S\'{e}rsic model does not always capture irregular features of the disk. 


\bibliography{article}{}
\bibliographystyle{aasjournalv7}



\end{document}